\documentstyle[galley]{mn}
\begin{document}
\large
\baselineskip 18 truept
\pagestyle {plain}

\def \mnras {MNRAS}
\def \apj {ApJ}
\def \aap {A\&A}

\title [Comparison of Approximations]
{Gravitational instability in the strongly nonlinear regime:
A study of various approximations}

\author [B.S. Sathyaprakash, et al]
{B.S. Sathyaprakash$^1$, V. Sahni$^1$, D. Munshi,$^1$ D. Pogosyan,$^2$
A. L. Melott$^3$\\
$^1$Inter-University Centre for Astronomy and Astrophysics
Post Bag 4, Ganeshkhind, Pune 411 007, India\\
$^2$Canadian Institute of Theoretical Astrophysics, University of Toronto,
Ontario, Canada\\
$^3$Department of Physics and Astronomy, University of Kansas, U.S.A.}
\maketitle

\begin {abstract}
We study the development of gravitational instability in the strongly
non-linear
regime. For this purpose we use a number of statistical indicators such as
filamentary statistics, spectrum of overdense/underdense regions and the void
probability function, each of which probes a particular aspect of gravitational
clustering. We use these statistical indicators to discriminate between
different approximations to gravitational instability which we test against
N-body simulations. The approximations which we test are, the truncated
Zel'dovich approximation (TZ), the adhesion model (AM), and the frozen flow
(FF)
and linear potential (LP) approximations. Of these we find that FF and LP break
down relatively early, soon after the non-linear length scale exceeds $R_*$ --
the mean distance between peaks of the gravitational potential. The reason for
this break down is easy to understand, particles in FF are constrained to
follow
the streamlines of the initial velocity field. Shell crossing is absent in this
case and structure gradually freezes as particles begin to collect near minima
of the gravitational potential. In LP particles follow the lines of force of
the
primordial potential, oscillating about its minima at late times when the
non-linear length scale $k_{\rm NL}^{-1}\simeq R_*$. Unlike FF and LP the
adhesion model (and to some extent TZ) continues to give accurate results even
at late times when $k_{\rm NL}^{-1} \ge R_*$. This is because both AM and TZ
use
the presence of long range modes in the gravitational potential to move
particles. Thus as long as the initial potential has sufficient long range
power
to initiate large scale coherent motions, TZ and AM will remain approximately
valid. In relation to AM, TZ suffers from a single major drawback -- it
underestimates the presence of small clumps.  Similarly, it predicts the right
mean density in large voids but misses subcondensations within them. The reason
for this is clear: The artificial removal of power on scales smaller than
$k_{\rm NL}^{-1}$ in the initial potential in TZ, designed to prevent shell
crossing, causes a substantial fraction of matter (which would have been
clustered in N-body simulations) to lie within low density regions at all
epochs.  On the other hand, TZ is very fast to implement and more accurately
predicts the location of large objects at late times.
\end {abstract}

\begin {keywords}
Cosmology : theory -- galaxies : formation -- large scale structure of
Universe -- methods : statistical
\end {keywords}

\section {Introduction}

The Universe on large scales exhibits remarkable structural features
as demonstrated by the numerous investigations of its statistical
properties.  It is believed that this structure arose via
amplification, through gravitational instability, of primordial
fluctuations in the density of matter. The evolution of such
fluctuations can be studied using the well known hydrodynamical
equations for the gravitating fluid.  In the past decade several
workers have obtained numerical solutions to these equations
which confirm that gravitational instability can lead to the
kind of structure observed in the Universe today.

While numerical N-body simulations are mandatory to approach the
precise picture, often our understanding of the dynamical
processes that lead to these structures comes from various
approximations to the fully nonlinear equations that have been
propounded. For instance, Zel'dovich showed that gravitational
instability generically leads to the formation of two-dimensional
sheets, the so called pancakes, the adhesion model which in some sense
can be regarded as an extension of the Zel'dovich approximation,
demonstrated that matter moves along pancakes towards filaments (which
form at the intersection of two pancakes) and then along filaments
towards clumps which form at the junction of two filaments (or three
pancakes). Thus the Zel'dovich approximation and the adhesion model showed that
gravitational instability leads to
the formation of cellular structure described by pancakes, filaments and
clumps -- a result that has also been independently verified by detailed
N-body simulations \cite{z70,messks83,ms89,sz89,mps94,sc94}.
In recent years several other approximations to the
nonlinear equations governing gravitational instability have been proposed.
Such approximation schemes serve a dual purpose:
Firstly, they have the potential to provide us with insight
regarding the physical processes which led to the formation
of structure. Secondly, they are as a rule easier to
implement and are often computationally less expensive than full N-body
simulations.
In order to apply a given approximation effectively
we should have a clear understanding of the domain of its validity.
It might also so happen that
certain statistical properties are reproduced by an
approximation to the same level of accuracy as in an N-body simulation
although certain other statistical properties may be reproduced to a
much lower accuracy.
For instance, a given approximation might correctly reproduce the void
probability function at a given epoch and yet fail to give the correct
multiplicity function for overdense regions.
It is therefore essential to examine different non-linear approximations
with a number of distinct (and in some cases orthogonal) statistical
discriminators.

In the present paper we compare the following approximation methods both
with each other and with the results of N-body simulations performed
using a two--dimensional PM code running with 512$^2$ particles on a 512$^2$
mesh (for details see \cite{bdmps91}). We tested:
(a) Zel'dovich approximation (truncated version),
(b) the adhesion model,
(c) frozen flow approximation and
(d) linear potential approximation.
In a parallel study, we follow the development of gravitational instability
using a variety of statistical indicators which probe its different features.
The statistical indicators which are used for comparing (a) -- (d)
single out certain features of non-linear clustering such as the existence
of voids, filaments (analogues of pancakes in 2D) and clumps.
The domain of validity of a given approximation is therefore discussed
with reference to a given statistical indicator.

The treatment followed in this paper extends earlier work
of Coles, Melott \& Shandarin (1993) in
which three nonlinear approximations:
the Zel'dovich approximation, the truncated Zel'dovich approximation, and
the lognormal approximation were compared with N-body simulations.
The present treatment is also in a sense complementary to recent work
\cite{ms93,bsbc93,mss94},
in which several of the approximations considered by
us were examined in the weakly
nonlinear regime of gravitational instability.
A common conclusion drawn in the above papers was that the Zel'dovich
approximation was more accurate than either the frozen flow or the linear
potential approximation when tested against the results of perturbation
theory in the quasi-linear regime. (This analysis was generalised to
include Lagrangian perturbation theories in Munshi, Sahni \& Starobinsky
(1994). The present paper
presents a fully nonlinear treatment of the problem thereby
considerably extending the quasi-linear analysis of the above authors.
Our tests were conducted in two dimensions and complement the three
dimensional analysis of Melott et al. (1994) and
Melott, Shandarin \& Weinberg (1994).
Some of our results may however
be carried over to three dimensions as well.
The exact formulation of the truncated Zel'dovich approximation used here
may be found in Melott, Pellman \& Shandarin (1994).

The paper is organised as follows.
In section II we briefly discuss the various approximations
to gravitational instability that we have chosen to compare with
N-body simulations. Section III is divided into several
subsections each of which deals with a particular statistic.
In section IV we summarize the chief results of our investigations.

\section {Nonlinear Approximations}

Consider pressureless matter with density $\rho(t,{\bf x}).$
The dynamics of such a fluid is governed
by the expansion of the Universe as also
by inhomogeneities in its distribution.
The component of the velocity
which arises solely due to inhomogeneities in the density field
is known as the peculiar
velocity ${\bf v}(t,{\bf x}) = \dot{\bf r} - H\bf r$.
The combined evolution of the density $\rho$, peculiar velocity and the
peculiar gravitational potential
$\varphi (t,{\bf x})$ is given by the following well
known system of coupled nonlinear equations:
\begin{equation} {\partial \rho \over \partial t} +
3H \rho +
{1\over a} \nabla \cdot (\rho {\bf v}) = 0,
\label {conserve} \end{equation}

\begin{equation} {\partial {\bf v}\over \partial t} +
{1\over a} ({\bf v}\cdot  \nabla) {\bf v} +
H {\bf v} =
-{1\over a} \nabla \varphi,
\label {euler1} \end{equation}

\begin{equation} \nabla^2\varphi = 4 \pi G a^2 (\rho-\rho_0),
\label {poisson} \end{equation}
where $a(t)$ is the cosmic expansion factor, $H$ is the Hubble
parameter and $\rho_0$ is the average density of the fluid.
The spatial derivatives in (\ref{conserve}) -- (\ref{poisson})
are defined with respect to the
comoving coordinate ${\bf x} = {\bf r}/a$.
Choosing a new time variable $a(t)$, and defining a {\it comoving}
velocity variable
\begin{equation} {\bf u} = {d{\bf x} \over da} = {{\bf v} \over a \dot a},
\label {covel} \end{equation}
we obtain the following form for the Euler equation (\ref {euler1}):

\begin{equation} {\partial {\bf u} \over \partial a} +
({\bf u} \cdot \nabla) {\bf u} =
- {3\over 2 a} ({\bf u} + A \nabla \varphi),
\label {euler2} \end{equation}
where $A=2/(3H^2a^3)$ is a constant for a flat Universe with dust-like
matter. In cosmological problems, where initial perturbations correspond to
the growing scalar mode, the velocity field $ {\bf u} $ is potential
$ {\bf u} = - \nabla \Phi $ until multistream regions develop.

At earlier moments of time when inhomogeneities are small the solutions to
equations (\ref{conserve})-(\ref{euler2}) may be obtained by linearization.
We then have during the linear stage
\begin{equation}
{\bf u} ({\bf q}) = -A \nabla \varphi ({\bf q})
\label{linear}
\end{equation}
where ${\bf q} $ are the initial (Lagrangian) coordinates. Therefore, initally,
the velocity potential and the gravitational potential are simply proportional
to one-another
$ \Phi ({\bf q}) = A \varphi ({\bf q}) $, both unchanging in a
flat matter dominated Universe.

Later, in the nonlinear regime, there is no easy solution to the basic system
(\ref{conserve})-(\ref {euler2}). The
different nonlinear approximations considered by us can be
conveniently described as
different ways of simplifying equation
(\ref {euler2}).

\subsection {Truncated Zel'dovich Approximation (TZ)} \label {tza}

The Zel'dovich approximation (henceforth ZA)
may be obtained from (\ref {euler2}) by
setting its right hand side to zero:
\begin{equation} {D {\bf u} \over D a} \equiv
{\partial {\bf u} \over \partial a} +
({\bf u} \cdot \nabla) {\bf u} = 0.
\label {zeldovich0} \end{equation}
where $D/Da$ is the convective derivative.
The above equation says that the dynamics of the
fluid element is governed  purely by ``inertia''.
It has an immediate solution in terms of the displacement of the fluid element
from its initial position with constant velocity \cite{z70}
\begin{equation}
{\bf x} = {\bf q} + a(t) {\bf u} ({\bf q})
\label {zeldovich}
\end{equation}

By setting the right-hand side of eq.(\ref{euler2}) to zero in ZA, one
extrapolates the linear relation (\ref{linear}) between velocity and
gravitational potential into nonlinear regime where the potentials are
generally time dependent ${\bf u} ({\bf x},t)= -A \nabla \varphi ({\bf x},t)$.

ZA works reasonably well so long as
streamlines of flows do not cross one another.
However, multistream flows invariably form at the locations of pancakes,
which grow progressively thicker leading to the ultimate
break down of the Zel'dovich approximation \cite{sz89}.
An extension of ZA called the truncated Zel'dovich approximation
\cite{cms93} is based on the
observation that
the formation and thickening of pancakes can be delayed
by artificially removing power on all scales smaller than the
one that is currently going nonlinear.
The length scale and window shape with
which the original spectrum is best smoothed has been determined in
three dimensions by Melott, Pellman \& Shandarin (1994).
In our simulations we have used
a $k$-space Gaussian window $\exp(-k^2/2k^2_G)$ to implement the necessary
truncation. As found by Melott, Pellman \& Shandarin (1994).
the optimal cutoff scale is
related to the scale entering nonlinearity. However, the precise
filtering scale $k_G^{-1}$ depends on the spectrum. This represents a
drawback of the model as we cannot obtain the best results for an arbitrary
spectrum based on first principles.  However, the cutoff scale is only
weakly spectrum dependent.  The optimal value
of $k_G^{-1}$ for spectra considered in this paper will be
discussed in the next section (cf. Table II). Nevertherless, let
us note that even if we do not use the best filter for a given spectrum but
a fixed spectrum independent filtering $k_G \ne k_G(n)$ the approximation
still retains most of its positive features. One major advantage of the
TZ is the extreme simplicity of its implementation.

An extension of TZ is the use of second order perturbation theory
combined with the smoothing of the initial potential.  This produces
somewhat more accurate but not qualitatively different results from
TZ \cite{bmw94,mbw94}.  We do not include this second-order approach
in this study.
\subsection {Adhesion Model (AM)} \label {sam}

The adhesion model is an extension of the Zel'dovich approximation.
In the adhesion approximation the right
hand side of (\ref {euler2}) is replaced by an artificial
viscosity term to mimic the effects of nonlinear gravity on small
scales and to stabilise the thickness of pancakes.
The resulting equation is the well known Burger's equation and has the
form \cite{b74,gss85}
\begin{equation}
{\partial {\bf u} \over \partial a} +
({\bf u} .\nabla) {\bf u} = \nu \nabla^2 {\bf u},
\label {am} \end{equation}
where $\nu$ is the coefficient of viscosity.
It is interesting that
in the limit $\nu \longrightarrow 0$ the right hand side of
(\ref {am}) remains finite
only in those regions where large gradients in the velocity field
exist (viz inside the pancakes), and vanishes elsewhere.
As a result the adhesion model reproduces the results of the Zel'dovich
approximation exactly in regions outside of the pancakes themselves.
Accordingly, the adhesion model reduces to ZA for the early time moments
or sufficiently smoothed initial conditions when no shell-crossing
is present.
For vanishing $\nu$ the adhesion model has an elegant geometrical
interpretation which we have used to construct the skeleton of the
large scale structure predicted by this model
\cite{gss85,pog89,kps90,kpsm92,sss94}.
An undesirable
limitation of the geometrical prescription is that it
does not give particle positions but only
locations of filaments and clumps which have to be smoothed by an
appropriate filter in order to lend themselves to a comparative
treatment with other models and with N-body simulations.
A study using particles \cite{msw94} in three dimensions shows general
agreement with our results when equivalent tests were done.

\subsection {Frozen Flow Approximation (FF)} \label {sff}

The underlying philosophy of FF is in a sense just the converse of ZA
since the inertia of particles is neglected in this approximation which
requires particles to constantly upgrade their velocity to a value
determined by the local value of the linear velocity field.
More precisely, FF corresponds to neglecting
both the nonlinear term, namely, $({\bf u}\cdot \nabla) {\bf u}$
and the right-hand side in (\ref {euler2}) \cite {mlms92}
\begin{equation} {\partial {\bf u} \over \partial a} = 0
\label {ff} \end{equation}
so that the comoving velocity field remains fixed to its linear value
${\bf u} ({\bf x},t) = {\bf u} ({\bf q} ={\bf x}) $.
It is clear that matter in FF is collected with time in the points
$ {\bf u} ({\bf q}) = 0 $ i.e. in the positions of the local minima of the
initial gravitational potential. Therefore FF cannot be expected to work
even qualitatively for late time moments when the scale of nonlinearity
escalates above the typical distance between minima of the initial
potential.

\subsection {Linear Potential Approximation (LP)} \label {slp}

N-body simulations show that the gravitational potential evolves
much more slowly than the density field \cite {bsv93}. This is so
because relative
to $\delta$ the potential $\varphi$ is
dominated by small--$k$ modes which obey the precepts of linear theory
longer than large--$k$ modes.
The linearized Poisson equation $\nabla^2\varphi = 4\pi G a^2
\delta\rho$ demonstrates that $\varphi \simeq {\rm const.}$ as long as
$\delta \ll 1$ and the Universe is flat and matter dominated.
Extending this assumption ($\varphi \simeq {\rm const.}$)
into the nonlinear regime as well, we arrive at the following
generalisation of the Euler equation which describes the dynamics of the
linear potential (or {\it frozen} potential) approximation \cite{bsv93,bp94}
\begin{equation} {\partial {\bf u} \over \partial a} + ({\bf u} \cdot \nabla)
{\bf u} = - {3\over 2 a} ({\bf u} + A \nabla \varphi_0),
\label {lp} \end{equation}
where $\varphi_0 \equiv \varphi ({\bf x}, t_0) = \varphi ({\bf q}) $.
This equation defines the force acting on a fluid element
at the instant $a(t)$ using the primordial value of the potential
$\varphi = \varphi_0.$
In a sense the LP can be regarded as an N-body simulation in which the
value of the potential is not upgraded after each time step.

Both TZ and AM have single-step analytical solutions.
Consequently, there is no need to
evolve the fluid iteratively; given some initial conditions these
approximations have the ability to directly give the configuration at any
epoch which may be of
interest. In contrast,  LP and FF are Eulerian approximations
and have no analytical solutions except in some
special cases. Operationally they are similar to full N-body simulations
which evolve the fluid iteratively, except for the fact that in LP the
potential
is kept frozen to its initial value and in FF
neither is the velocity potential upgraded nor is particle inertia taken into
account.
Since PM type N--body simulations are easy to do on modern computers, it is not
clear whether LP and FF have value beyond the descriptive insight which they
provide.

\section {Comparative study of various approximations}

In this section we employ a number of statistical tools to
compare the approximations mentioned in the previous section
with N-body simulations. We use the same initial conditions
for all the approximations and they form a subset of the initial conditions
used by Beacom et al. (1991) and Kofman et al.
(1992) for other purposes. Time evolution
of the N--body models can be seen in the video accompanying Kofman et al.
(1992).
All our comparisons
are carried out in two dimensions with the initial
potential being specified on a grid of size $512 \times 512.$
More specifically,
the models for which we have carried out the comparison are either
featureless or truncated power law spectra of the general form
\begin{equation}
P(k) \propto \left \{
\begin{array}{l}k^n, \ {\rm for}\ k \le k_c; \\ 0,   \ {\rm for}\ k>k_c.
\end{array} \right.
\end{equation}
We have considered three different spectral indices, $n=2,~0,~-2,$
with a cutoff in each case, at the Nyquist wavenumber:
$k_c= 256 k_f,$ where $k_f$ is the fundamental mode.
In addition to this, we have a $n=0$ model with a truncation
at $k_c=32 k_f$ which serves to illustrate the effect of
an abrupt cutoff in the power spectrum as happens in some
models of dark matter like hot dark matter.
Thus, we have a total of four models in all.

All our simulations of the various approximations are performed
using a particle code excepting the adhesion model which
is simulated using the well known geometrical interpretation
of the solution to Burger's equation
\cite{sz89,pog89,sss94}.
Consequently, we could not include adhesion in studying some
statistical properties, such as the position correlation coefficient or
filament statistics, which rely on having
particle positions. Where we could compare AM with N-body we expect
the former to do somewhat better than what our results convey.
See also \cite{msw94} for a
particle--based three--dimensional study of AM.

We compare the evolved density fields and the quantities derived
from them when different scales are going nonlinear. We choose
$\sigma(k_{\rm NL}),$ the epoch when the scale
$2\pi/k_{\rm NL}$ is going nonlinear,
as a convenient measure of ``time'' with which to characterize different
regimes in nonlinear gravitational clustering:
\begin{equation}
\sigma(k_{\rm NL}) =
\left ( {{ \int_{k_f}^{k_{\rm N}} {P(k) k dk}\over
\int_{k_f}^{k_{\rm NL}} {P(k) k dk}}}
\right )^{1/2}.
\label {epochdef} \end{equation}
Here $k_{\rm N}$ is either the Nyquist wavenumber or the cutoff mode $k_c,$
whichever is smaller. For truncated power law spectra (with a cutoff at $k_c$)
\begin{equation}
\sigma (k_{\rm NL}) = \left ({k_c \over k_{\rm NL}} \right ) ^{{n+2}\over 2},
\quad n\ne -2,
\label {epoch1} \end{equation}
and
\begin{equation}
\sigma (k_{\rm NL}) = \left ( {\ln \left (k_c/k_f \right ) \over
\ln \left (k_{\rm NL}/k_f \right )} \right ) ^{1\over 2},
\quad n=-2.
\label {epoch2} \end{equation}
The first scale to go nonlinear
is the one corresponding to either the Nyquist wavenumber (in the case of
models with no cutoff) or the mode $k_c.$ When this
happens, by definition, $\sigma=1.$ As $\sigma$
increases, successively larger scales enter the nonlinear regime.
For concreteness we have considered in our comparison those
values of $\sigma$ for which the scales going nonlinear are
$k_c,$ $k_c/2,$ etc., and we stop the integration when the
scale entering the nonlinear regime comes close to the size of the
simulation box.

In this study we suggest that two natural scales characterizing
a given model may be well suited for giving bounds on the validity
of some approximation methods.
These are: (i) the scale $R_*,$ corresponding
to the average distance between the peaks of the potential
and (ii) the scale $R_\varphi$ characterizing the correlation length
of the potential.  They are given in terms of the moments of the
potential field by the following expressions:
\begin{equation}
R_\varphi = \sqrt 2 {\sigma_0 \over \sigma_1}
\quad \quad
R_* = \sqrt 2 {\sigma_1 \over \sigma_2}
\label {scales} \end{equation}
where the moments $\sigma_j$ are defined by
\begin{equation}
\sigma_j^2 \propto \int_{k_f}^{k_c} k^{2j-4}P(k) k dk.
\label {moments} \end{equation}
The epoch $\sigma_*$ ($\sigma_\varphi$) when the scale
$R_*$ ($R_\varphi$) is going nonlinear
can be found from (\ref {epoch1}) and (\ref {epoch2}) by substituting
$k_{\rm NL}=R_*^{-1}$ $(R_\varphi^{-1}).$
The values of $\sigma_*$ and $\sigma_\varphi$ are listed
in Table I for the various models under discussion.
The values of $R_*$ and $R_\varphi$ as well as $\gamma = R_*/R_\varphi$
are plotted in Fig. \ref {gamma} as functions of the spectral index $n$.

\begin{table*} \begin{center} \begin{tabular}{ccccc} \hline \hline

& \multicolumn{2}{c}{$k_c=32 k_f$} & \multicolumn{2}{c}{$k_c=256 k_f$}\\ \hline
n & $\sigma_*$ & $\sigma_\varphi$
  & $\sigma_*$ & $\sigma_\varphi$
\\ \hline
$-2$  & 2.36 & $\infty$ & 2.55  & $\infty$ \\
\ \ 0 & 3.75 & 20.3     & 4.71  & 120      \\
$+2$  & 4.00 & 14.0     & 4.00  & 22.2    \\
\hline \end{tabular} \end{center}

\caption{The scales $R_*$ and $R_\varphi$
of potential corresponding to the average distance between the
peaks of the potential and the correlation length, respectively.
The box is assumed to be of unit length.
Also tabulated are the corresponding epochs $\sigma_*$ and $\sigma_\varphi$
when these scales go nonlinear.} \label{tabscales} \end{table*}

The values of $R_*$ and $R_\varphi$ as well as $\gamma = R_*/R_\varphi$
are plotted in Fig. \ref {gamma} as functions of the spectral index $n$.

We recognize that the specific values of $\sigma_*(\sigma_\varphi)$ and
$R_*(R_\varphi)$
are often (and in particular in our case of power-law
initial potentials) determined mainly by numerical cutoffs
introduced by the limitations of our computer simulation.
In fact this reflects once again
the effect of the finite grid used in any simulations
on the representation of the underlying initial spectrum.
For the real Universe and spectra such as CDM, physical cutoffs
are provided by the horizon scale (for~$R_\varphi$) and by the
free-streaming distance (for $R_*$).

As mentioned earlier in Sec. \ref {tza} the optimal smoothing scale
that needs to be used in TZ simulations depends on the spectrum.
By definition, the optimal smoothing scale is that scale which obtains
the maximum correlation coefficient of TZ density fields with
N-body density fields. We have found that its relation to
the scale entering nonlinearity is fairly independent of the epoch.
(If this were not so then the very concept of truncated Zel'dovich
approximation would lose its meaning.)
Table II lists the optimal smoothing scales $k_G^{-1} = k_{\rm opt}^{-1}$ for
different spectra considered by us.

\begin{table*} \begin{center} \begin{tabular}{ccc} \hline \hline

n & $k_{\rm opt}/k_{\rm NL}$ \\
$-2$  & 0.5---1.50 \\
\ \ 0 & 1.25 \\
$+2$  & 1.00 \\
\hline \end{tabular} \end{center}

\caption{The optimal smoothing scale $k_{\rm opt}^{-1}$ used in the
truncated Zel'dovich approximation scheme depends on the index of
the power spectrum as shown above. For $n=-2$ spectra the density
correlation coefficient is virtually the same for a wide
range of values of $k_{\rm opt}$.} \label{optscales} \end{table*}

\subsection {Visual comparison} \label {vc}

To begin with, we make a visual comparison of the various
approximation schemes with N-body simulations. The structure
obtained using the AM and the evolved particle
positions in the case of FF, LP and TZ are shown in
Fig. \ref {visual}a-d for four different spectra.
 From top to bottom, the pictures correspond to
N-body, adhesion model, frozen-flow approximation,
linear potential approximation, and truncated
Zel'dovich approximation, respectively.
Fig. \ref {visual}a corresponds to $n=0,$ $k_c=32 k_f$ model and
Fig. \ref {visual}b to $n=0,$ $k_c=256 k_f$ model.
In Fig. \ref {visual}c and \ref {visual}d the spectral index is
$n=2,$ and $-2,$ respectively, and $k_c= 256 k_f$ in both the cases.
In Fig. \ref {visual}a, b and d
the left hand panels correspond to an epoch $\sigma$ such that
$\sigma \sim \sigma_*$ $(k_{\rm NL}^{-1} \sim R_*)$ and the right
panels to an epoch $\sigma \sim \sigma_\varphi$
$(k_{\rm NL}^{-1} \sim R_{\varphi}).$
In Fig. \ref {visual}c both the left and the right hand panels correspond to
an epoch $\sigma>\sigma_{\varphi}$. (In the case of $n=2$ models
there is a lot of small scale power. Consequently, the pictures
at a stage when $R_*$ is going nonlinear looks too grainy.
Therefore it is not easy to compare them visually at that
epoch.)

We find that at the epoch when the scale going nonlinear
is $R_*,$ and at earlier epochs, all the approximation
schemes appear to reproduce the structure with roughly the same accuracy
as in N-body simulations. (This is also reflected by the
high value of the correlation coefficient before the epoch $\sigma_*$
as discussed in sections \ref {ccp} and \ref {ccd} -- see Fig. \ref {vccfig}
and \ref {sccfig}).

The epoch corresponding to $k^{-1}_{\rm NL} \simeq R_*$ characterizes
the completion of cellular structure which forms from an initially smooth
distribution of matter.
Later epochs are characterised by the relative motion and mergers of the
structure elements governed by mutual
attraction of large mass concentrations (knots and filaments) as well as
repulsion from underdense interiors of cells (voids). It is clear that
both FF and LP which fix the structure on scale $ R_*$ are unable
to describe this process even qualitatively, and therefore begin to fail
beyond the epoch $\sigma_*.$ These conclusions are borne out by
Fig. \ref {visual}a-d.

 From the right panels it is clear that close to the epoch when the
scale going nonlinear is $R_\varphi$ the structure obtained using
AM is still in excellent visual agreement with that of N-body simulations.
TZ has a reasonably good visual agreement with N-body on large scales
though at this epoch the small scale features abundant
in N-body simulations (especially for spectra with $n\ge 0$)
are not present in the TZ simulation. However, much
before this epoch, in fact soon after the scale $R_*$ has crossed
nonlinearity, FF and LP approximations fail to give the right
picture. Particles in FF, approach the valleys and the minima
of the potential asymptotically, leading to greatly thinned out
filaments which eventually empty out and vanish as the
particles gradually fall into the minima of the potential. In the case of
LP, particles execute oscillations around
the troughs of the primordial potential partially simulating, at earlier
epochs, the results of N-body simulations wherein the pancakes neither
thicken as in the Zel'dovich approximation nor do they thin out as in FF.
The contour diagrams of the initial potentials used in our
simulations, shown in Fig. \ref {contour}, bear out this claim about the
behaviour of particles in FF and LP.
Notice especially the pictures corresponding to the $n=2$ power-law
model (cf. Fig. \ref {visual}c)
which has a lot of small scale power. Here we see that the
particle positions in FF and LP are essentially frozen beyond the epoch
$k_{\rm NL}^{-1} \simeq R_*.$
While the dynamics in LP and FF at all times are determined by the
gradients of the local primordial potential, we know from N-body
simulations that beyond the epoch of formation of cellular structure,
the small scale features of the primordial potential play little, if any,
role in the furtherance of gravitational
clustering \cite {bdmps91,pogthe,weinb,pm94}.
It is therefore to be expected
that LP and FF will not be able to reproduce qualitatively the features of
hierarchical clustering.

The visual agreement of the pictures obtained using AM and TZ with N-body
lasts for epochs
much longer than that for either FF or LP since the former two approximations
successively use power on larger scales to influence
the dynamics of the fluid. Even though the local gravitational
potential has changed substantially by the time $R_*$ has gone nonlinear,
it has not evolved so much as to compete with the effect of
power on large scales. Thus, as long as the initial potential
has sufficient large scale power to give rise to coherent motion over
large scales, TZ and AM will remain approximately valid. Following
Kofman et al. (1992)
we speculate that the coherence length $R_\varphi$
of the primordial potential is important in this discussion.
By the time $\sigma _{\varphi}$ when the scale $R_\varphi$
goes nonlinear, in fact even at slightly earlier epochs,
both AM and TZ begin to produce structure substantially different in small
detail from that seen in N-body simulations.
However, it is impossible to determine from our results whether this change
is due to a transition at $R_\varphi$ or our increasing ability to resolve
detail as the simulation goes nonlinear on larger scales.
TZ does not produce small objects, and AM
puts them in the wrong place.
Pauls \& Melott (1994) have shown that even at
much later times than $\sigma _{\varphi}$,
the primordial potential
can determine the coherent motion of large clumps, and TZ can produce correct
positions for them while AM begins to make errors in the position of large
objects as well and all other approximations have broken down long before.
However, let us stress that both TZ and AM continue to
reproduce qualitative features of the structure even beyond
$R_\varphi$.

The TZ approximation suffers from
one major drawback. Since in this approximation we have artificially removed
power on small scales, matter never gets collected in small clumps.
It does, however, put about the right amount of mass in large clumps.
Consequently,
a substantial fraction of matter (which would be in small clumps in an
N--body simulation)
lies within low density regions at all epochs. As a result TZ (which has the
advantage of being computationally very fast) is not well
suited for studying
small clumps even though it gives a remarkably good
correlation coefficient when compared with N-body simulations.
Similarly TZ gives the right mean density in large voids but
misses subcondensations within them.
This is the price it pays for its greater accuracy in locating the large--scale
mass distribution.
These views, based largely on a visual comparison of the different
approximations, are borne out by more quantitative comparisons which we
discuss below. The AM suffers from two major drawbacks. First,
it is computationally expensive, sometimes
approaching the cost of an N--body simulation.
Second, although it does produce small clumps, it occasionally makes major
errors in their
position (sometimes comparable with the scale of nonlinearity),
especially at late times and larger $n$.

\subsection {Quantitative comparison} \label {qc}

We now turn to a quantitative comparison of the various approximation
schemes with N-body simulations.
The information about any statistic can in principle be inferred
from a knowledge of all the N-point correlations, or at least a
substantial number of them. In practice, however, the full hierarchy of
correlation functions is virtually impossible to calculate
due to heavy computing requirements. Even if it were in
principle possible to compute the higher order correlation functions
it is not clear how much light that would shed on
structural units that are often of interest such as clumps and filaments.
It is with the aim of studying such structural units that we have chosen
a number of statistical tools each of which addresses a specific
structural feature present in the simulation.
To this end we choose the time evolution of
the following indicators for comparison:

\begin{enumerate}

\item Correlation coefficient of particle positions
of different approximation schemes with N-body,
\item Correlation coefficient of density fields of different
approximation schemes with N-body,
\item Number of {\it clumps,} (defined as regions of a certain overdensity),
\item {\it Filament} statistics,
\item Number of {\it voids,} (defined as regions of a certain underdensity),

and

\item {\it Void probability function.}
\end{enumerate}

We note that the first two indicators are primarily dynamical, in that they
test for specific point--by--point agreement between the approximation and the
N--body simulation. The others are global statistics, and test for particular
kinds of similarity.

\subsubsection {Correlation coefficient of particle positions} \label {ccp}

Given an approximation scheme an obvious question that comes to mind
is how well can the approximation scheme evolve the particles
in relation to the exact N-body simulations. To answer this question
we consider the correlation coefficient of the coordinate positions
of an approximation scheme and N-body. Let ${\bf X}_{\rm A}^i (\sigma)$
and ${\bf X}_{\rm N}^i (\sigma)$ denote the position of the $i$~th
particle, at an epoch $\sigma (t),$ in an approximation A and N-body,
respectively. The displacement of the particles from their unperturbed
coordinates ${\bf X} (\sigma (t_0))$ is given by:
\begin{equation}
\Delta {\bf X} (\sigma (t)) = {\bf X} (\sigma (t)) -
{\bf X} (\sigma (t_0))
\label {diffposition} \end{equation}
The linear correlation coefficient of two vector fields
$\Delta {\bf X}_{\rm A} (\sigma (t))$ and
$\Delta {\bf X}_{\rm N} (\sigma (t))$
is defined by
\begin{equation}
r_{{\bf X}} \equiv
{
\sum_i\delta {\bf X}_{\rm A}^i \cdot \delta {\bf X}_{\rm N}^i
\over
\left [
\sum_j(\delta {\bf X}_{\rm A}^j )^2
\sum_k(\delta {\bf X}_{\rm N}^k )^2
\right ]^{1/2}
}
\label {vcc} \end{equation}
where a dot denotes the scalar product of the two vector fields,
where $\delta {\bf X} \equiv \Delta {\bf X} - \left <\Delta {\bf X}\right>$
is the deviation of the displacement vector from average displacement,
where a summation is over the entire sample
and where $\left <~\right >$ indicates average over the entire sample.
This is a straightforward generalization of the familiar correlation
coefficient defined for scalar fields.
In Fig. \ref {vccfig} we have plotted the evolution
of $r_{{\bf X}}$ for different approximation
schemes, and for different spectra, as a function of $\sigma (t).$
Here, and in Fig. \ref {vccfig} and \ref {voids}, we adopt the
following scheme for displaying the evolution of different statistics:
The top left panel corresponds to $n=0,$ $k_c=32 k_f$ model
and the top right panel to $n=0,$ $k_c= 256 k_f$ model. The
bottom left panel corresponds to $n=2$ model and the bottom right
panel to $n=-2.$
In Fig. \ref {vccfig} the results AM are missing
since our implementation of this approximation scheme (using an osculating
paraboloid) does
not obtain particle positions.

We observe that when the spectral index $n=-2$ there is
an excellent agreement between all the three approximations
and the N-body. ($r_{\Delta {\bf X}}$ is always
larger than about 0.9 for all approximations for this spectrum.)
In the $n=2$
case, surprisingly, the correlation vanishes to begin with,
building up later to reach a maximum
of about 0.6 before dropping. We do not understand
this behaviour entirely but suspect that some numerical effect from the
excessively large
power on very small scales present in this case
is the root cause for such a behaviour.
In this case we see that none of the approximations produce the
right kind of displacement of particles. This is absolutely in
agreement with Fig. \ref {visual}c (corresponding to $n=2$ spectrum)
wherein we see that in the case of both FF and LP matter
simply gets collected into the local minima of the potential
without ever transferring the power to larger scales while in the
case of TZ matter does not cluster on small scales at all.
Although the agreement of all approximations with N-body is relatively poor for
this spectrum,
we find that TZ gives consistently higher values for the correlation
coefficient especially at late
times when clustering is more prominent on large scales.

 From Fig. \ref {vccfig} we find that for spectra with
substantial power on a wide range of scale,
such as $n=0$ or $n=2$, particle positions in TZ agree with those in
N-body much better than do either LP or FF. This is because
of the fact that for such spectra, wherein both small and
large scales dictate the dynamics (with larger scales being more
important at later epochs) LP and FF break down at relatively
earlier epochs than TZ and AM.

\subsubsection {Correlation coefficient of density fields} \label {ccd}

While the correlation coefficient of particle positions tells
us precisely how the structure is produced in an approximation
scheme it is seldom the main quantity of interest; it is only
a measure of how good a dynamical approximation scheme is in relation
to the N-body solutions. One is often interested in
the evolution of the density field since it lets us
infer the evolution of many other structural
units such as clumps, filaments and voids. In order to study
the time evolution of the density field we obtain the
density field by employing the cloud-in-cell (CIC) algorithm.
This algorithm can only be used when particle positions
are known and hence cannot be directly used for our AM simulations. In
the latter case, we use the structural units of clumps
and filaments (plus the ``free'' particles -- those that
have not yet fallen into caustics ) given by the model to
reconstruct the density field. The density field for AM
simulations is obtained in three steps:
(i) The mass in the ``free'' particles is distributed using the CIC algorithm.
(ii) Each clump given by AM is assumed to be a ``Gaussian
hill'' with the variance of the Gaussian chosen to be proportional
to its mass.
(iii) The rest of the
mass is distributed uniformly amongst filaments which is then
smeared using the CIC algorithm.

The density fields so obtained are smoothed at a
certain scale before computing the correlation coefficient.
Such a smoothing is motivated by the fact that the density
fields evolved by approximation schemes
are not expected to agree with those of N-body simulations
in great detail; any agreement is to be expected only after the
small scale inhomogeneities are smoothed out. In fact, the
relevant question here is: ``How well can a given approximation mimic
the results of exact equations on medium scales?''.
The physical reason for this is that the very large scale properties of the
universe can be studied by ZA or Eulerian perturbation theory, while the
smallest ones by N--body simulations plus hydrodynamics.
We might mention that an improvement
in the correlation coefficient for TZ is expected if instead of
filtering the density on a fixed scale, say, $k \simeq 64k_f$, a variable
filter scale which is a constant multiple of the scale of non-linearity
$k \propto k_{\rm NL}$ is chosen. This is demonstrated in Fig.\ref{sccfig}b
for a density which is filtered on a scale $k=2k_{\rm NL}$ for the
spectrum $n=0,$ $k_c=256 k_f.$

Following Coles Melott \& Shandarin (1993)
we use the correlation coefficient
of density fields to compare the approximation schemes with N-body
simulations. Given density fields $\rho_{\rm A}(\sigma)$ and
$\rho_{\rm N} (\sigma),$ corresponding to an approximation scheme A
and the N-body simulation, respectively, the correlation coefficient of
these fields is defined by a formula similar to equation (\ref {vcc}):
\begin{equation}
r_\delta \equiv {
\sum_i \delta_{\rm A}^i \delta_{\rm N}^i
\over
\left [
\sum_j (\delta_{\rm A}^j )^2
\sum_k (\delta_{\rm N}^k )^2
\right ] ^{1/2}
}
\label {scc} \end{equation}
where $\delta\equiv (\rho - \rho_0)/\rho_0$ is the density contrast.

The evolution of the statistic $r_\delta$ is shown in Fig. \ref {sccfig}a.
The arrangement of the panels here is as in Fig. \ref {vccfig}.
Here we have also included the results of the adhesion model.
We notice that except for the $n=2$ model
TZ and AM are in better agreement with N-body than FF and LP.
For the $n=2$ model AM, FF and LP, all give roughly the same correlation
at all epochs, while TZ is better.
In contrast to the pictures wherein the agreement of TZ with N-body
is not so remarkable, the density correlation is extremely good
up to very late times.
In absolute terms however, both TZ and AM eventually break down.
The results of AM shown here are entirely consistent with those found
before \cite{msw94}
using a particle AM code, in spite of the artificial smoothing of
filaments and clumps that has gone into our code to make the geometrical
method mimic the particle method. In the case of $n=0$ power-law models,
we note that soon after the epoch $\sigma_*,$ say $\sigma=8,$
while the LP has managed to produce a correlation
coefficient of about 0.5, the FF has a far less value at this epoch.
However, both the AM and the TZ give values substantially larger than
0.5 indicating their validity at this and later epochs.
The reason why the FF does so badly can be traced to
Fig. \ref {visual}a-d wherein we see that matter has completely
emptied out into rivulets of the potential wells by the
epoch $\sigma_*.$

The superior performance of TZ on scales down to $k^{-1}_{\rm NL}$ combined
with
its speed makes it a good approximation for studying the mass
distribution from scales of about $b^{-1}(1+z)^{-1}10^{13} M_\odot$ on up,
where $b$ is the bias parameter and z is the redshift. As we shall see, it
fails on smaller scales. But the mass resolution is two orders of magnitude
better than in the past.

A good density correlation of an approximation scheme with an N-body
simulation does not necessarily mean that
other statistical indicators too will give good results when testing the
approximation with N-body (unless of course the density correlation approaches
unity).
Conversely, a poor density correlation
does not necessarily imply that the approximation will also predict
incorrect results for other statistics.
However, our two previous statistics, the
position and density correlations do test the dynamical accuracy of our
approximations. The tests which follow probe agreement with respect to
global statistical
measures, which is a somewhat different issue.
Correct global statistics do not necessarily imply correct dynamics.
In what follows, therefore,
we supplement the two correlation coefficients discussed above
with other statistical indicators that address issues relating to
the ensemble of structural units present in the simulation. They do not test
whether specific individual units are in same the same place
but they do test an overall  ``resemblance."

\subsubsection {Evolution of the number of clumps} \label {ec}

An important statistic which any theory of LSS
is expected to explain is the mass function of galaxies or of
clusters of galaxies: how the total mass is distributed in objects
of different masses. A related question is how the number of
clumps, defined as connected regions of a certain overdensity,
evolves with time. Due to lack of space, here we address only the
latter, relegating the more important former question to a future work.
The number of clumps, at any moment, clearly depends on the
density threshold $\rho_c$ we use to identify regions of overdensity.
However, since the aim here is to compare the predictions of
different approximations with N-body, it hardly matters what
density threshold we choose provided it suffices to obtain well
defined clumps. In Fig. \ref {overdensity} we have shown regions
in our N-body simulations of density $\rho \ge \rho_c$
at two different epochs each,
for the models $n=0,$ $k_c=32k_f,$ and $n=0,$ $k_c=256k_f.$
We see that the clumps are well defined
for the chosen density threshold and we use appropriate density
thresholds for the different spectra considered by us. A clump is now defined
as a connected region, in the sense of a ``friends-of-friends'' algorithm,
of overdensity greater than or equal to the density threshold $\rho_c.$

In Fig. \ref {clumps} we have shown the evolution of the number of clumps
for different simulations of the various power-law models discussed
earlier. The top panels show the evolution of the number of clumps
and the bottom panels show the evolution of the fraction of mass
in clumps.  In Fig. \ref {clumps}a
left panels correspond to $n=0,$ $k_c=32k_f$ model and the right
panels to $n=0,$ $k_c=256k_f$ model.
In Fig. \ref {clumps}b the left panels correspond to $n=2,$
and the right panels to $n=-2,$ models, respectively.
The results of N-body are shown in thick solid lines. Following the
N-body curve we see that generically, there are two distinct
phases in the clustering of matter via gravitational instability.
During the first phase the number of clumps keeps increasing
reaching a maximum after the epoch $\sigma_*$ at which time
the formation of cellular structure is complete. By this
epoch nearly 50 \% of the matter that ever gets bound
has fallen into the local wells of the
initial potential causing major changes in the local value of
the potential without, however, disrupting large scale modes.
During the second phase clustering proceeds hierarchically, with
smaller clumps merging with one another to form clumps of
larger mass. As a result the density contrast of the clumps
alone (not shown) keeps building up at a phenomenal rate
whereas the number of clumps begins to fall. Roughly, $\sigma_*$
characterizes the epoch of the transition from the cellular
to the hierarchical phase of clustering \cite{sss94}.
The gravitational potential on small scales undergoes
substantial changes during the second phase culminating
in the disruption of any initial small--scale coherence that might
have existed in the primordial potential.
Further discussion of the evolution of the potential can be found
in Pauls \& Melott (1994).
Beyond the epoch $\sigma_*$ no approximation scheme which proposes
to keep the local values of the potential unchanged, and at the
same time does not use large scale power in describing the dynamics,
can predict correct gravitational clustering.

Strictly speaking, as far the evolution of clumps is concerned
the agreement between N-body simulations and approximation schemes,
depends on the spectral index. However, based on Fig. \ref {clumps} we first
make the following general remarks. We note that with the
exception of AM none of the other approximations reproduce the
expected fall in the number of clumps at late times caused by
merger of clumps. In FF and LP the number of clumps, at earlier epochs,
show the general trend of a sharp rise and closely resemble the
predictions of N-body simulations till about $\sigma_*.$ However,
neither of these approximations show a proper fall off in
the number of clumps. In fact, with the exception of
the $n=0,$ $k_c=32 k_f$ model, there is no fall off seen in
the number of clumps in these approximations which is a major
feature of gravitational clustering:  In these approximations there is
relatively little evolution in the number of clumps
at later times. This is in contrast to N-body simulations wherein the merger
of clumps, with smaller clumps falling into the wells created
by the larger ones, is a never--ending process. The bottom
panels of Fig. \ref {clumps} lend further support to the viewpoint that
these two approximations do not predict the correct clustering of
matter beyond the epoch $\sigma_*.$ We observe that in N-body
simulations matter gets continuously drained into clumps whereas
in none of the approximation schemes, except in AM, is this
phenomenon seen generically.  We note that the statistics of
clumps produced by TZ is almost never in agreement with N-body
simulations. The reason for this is that in this approximation
the clumps are at no time well defined objects. Clumps
are relatively short scale features which form soon after
the rms linear density contrast reaches unity. Thereafter they
mature, and gain identity by the epoch when the scale going
nonlinear is $R_*.$ However, in the case of TZ the linear theory {\it rms}
density
contrast is never allowed to greatly exceed unity: power on successively
larger scales, in fact power on roughly the scale that is
entering nonlinearity, is filtered out. Consequently, although
voids are well defined in this approximation, clumps
never acquire a permanent identity.
The bottom panels of Fig. \ref {clumps}a show that hardly 50 \% of
the total mass ever gets collected in clumps in TZ. This, in
addition to the very low number of clumps that the approximation
predicts, makes the approximation scheme unsuitable for any
study concerning small scale features of LSS such as individual galaxies.
This is a shortcoming of this
approximation, but one that is not entirely unexpected.

In our opinion the AM is best suited in describing clustering
which statistically resembles N-body
simulations. It predicts the right kind of growth law for the fraction
of matter in clumps as well as the right merger histories of
collapsed objects (except perhaps in the $n=-2$ case).

\subsubsection {Filamentary statistics} \label {ef}

The second in our study of the structural units of LSS is
filaments. In order to understand the formation and
evolution of filaments we employ a statistic first
suggested by Vishniac (1979) and later used by Nusser and Dekel
(1990) to study filamentarity in different models of structure
formation. This statistic is obtained by first identifying the moments of
the distribution of particles around a chosen centre
and then constructing a scalar from these moments.

Let $M_k^\alpha (R)$ and $M_k^{\alpha\beta} (R)$ denote the
first and second ``moments'' of the distribution of the particles
located within a distance $R$ around the $k$~th particle:
\begin{equation}
M_k^\alpha = {1\over N_k} \sum_{j=1}^{N_k} x_j^\alpha,
\label {moment1} \end{equation}
\begin{equation}
M_k^{\alpha\beta} = {1\over N_k} \sum_{j=1}^{N_k} x_j^\alpha x_j^\beta.
\label {moment2} \end{equation}
Here $N_k$ is the number of particles within a distance $R$
from a centre located at the $k$~th particle, and $x_j^\alpha,$ $\alpha
= 1,2,$ denotes the
position vector of the $j$~th particle relative to the $k$~th
particle. The filamentary statistic $S(R)$ is the ensemble average
of the scalar $S_k(R)$ constructed for the chosen center
out of the two moments given above:
\begin{equation}
S_k(R) =
{2 \sum_{\alpha, \beta} M_k^{\alpha\beta} I_k^{\alpha\beta} -
 \sum_{\alpha}  M_k^{\alpha\alpha}
 \sum_{\beta} I_k^{\beta\beta}
\over
\left ( \sum_{\gamma} M_k^{\gamma\gamma} \right )^2},
\label {fs} \end{equation}
where
\begin{equation}
I^{\alpha\beta} =
M_k^{\alpha\beta} - M_k^\alpha M_k^\beta
\end{equation}
and
\begin{equation}
S(R) = {1\over N} \sum_{k=1}^N S_k(R)
\end{equation}
where $N$ denotes the number of centers chosen in carrying over the
average. The scalar $S_k(R)$ takes on values in the range $[0,1]$
attaining its maximum value of unity when the particles are aligned
along a straight line passing through the center and zero for a
uniform distribution of particles around the centre.

Evidently, using the above statistic we cannot infer about
filamentarity given only a density field since we must have
information about particle positions in order to compute
$S(R).$ Consequently, we will not be able to evaluate this
statistic for the adhesion model.  (It would certainly be
worthwhile to construct a statistic that would work for
densities.) For the rest of the approximations and N-body
we have chosen, at each epoch,  a random sample of about
2\%  of all the particle positions as ``test'' centers in
computing $S(R).$ The behaviour of $S(R)$ for N-body, FF, LP
and TZ simulations is shown in Fig. \ref {fil} for two
different epochs as indicated by the value of $\sigma$
quoted within the panels.  The two epochs chosen are exactly
as in Fig. \ref {visual} (see the caption of Fig. \ref {visual}).
The scale $R$ essentially characterizes the length,
in grid units, of the filaments being explored.

The N-body curves (thick solid lines) in Fig. \ref {fil} show
a clear transfer of power from smaller to larger filaments
as successively larger scales go nonlinear.
This behaviour, reflected by the evolution of the filamentary
statistic, is consistent with the visual impression rendered by
Fig. \ref {visual}a-d. For instance, the N-body panels on
the left in Fig. \ref {visual}a, which correspond to
earlier epochs, show predominantly small scale filamentarity
while those on the right, which correspond to later epochs,
show relatively large scale filamentarity.
On comparing the value of the statistic for different spectra
but at an epoch when the same scale is going nonlinear in all
of them we see that $S$ is larger for spectra with lower value of $n.$
Moreover, at later epochs there are in general two preferred
scales at which the filaments occur predominantly. This is
seen very clearly in the $n=-2$ model
(Fig. \ref {fil}b, bottom panels) wherein we see a lot
of power on very small scales, which sharply drops close
to zero on intermediate scales but rises again showing
substantial filamentarity on large scales. In short, the statistic
generically shows, at late times, two prominent peaks
(see right panels of Fig. \ref {fil}a and \ref {fil}b).
This is entirely consistent with what is expected  from
Fig. \ref {visual}a-d.
This demonstrates that the statistic $S$
truly characterizes filamentarity.
(One might want to increase the effectiveness of this statistic by
first using an algorithm such as the Minimal Spanning Tree to delineate
the main features of the distribution, and then apply the
filamentary statistic
exclusively to these features \cite {pc94,sc94}.)

At earlier epochs both FF and LP predict the right value of
the statistic on most scales and for all spectra. In the
case of $n=-2$ models there is an excellent agreement, at all
times, between FF, LP and N-body simulations. However, at later epochs
none of the approximations produce the right behaviour of the
statistic: FF predicts more power on smaller scales and lesser
power on larger scales while LP predicts lower power on almost
all scales. The reason why FF produces more power on smaller scales
can be seen from panels corresponding to FF in Fig. \ref {visual}a-d:
FF simulations show a lot of filamentarity on small scales and
none on large scales. Due to an inherent assumption
of this approximation there cannot be any shell crossing.
This means that particles approach the ``valleys'' of the
potential with ever decreasing speeds leading to many whisker
shaped objects. Matter that gets collected along the streamlines
of the velocity field in this way eventually drains out
into the minima of the potential wells. As a result, large
scale filaments never mature in this approximation. The problem
with FF is just the opposite to that with the Zel'dovich approximation:
In the latter the pancakes do not remain thin and in the former
the pancakes never thicken. Both these aspects are contrary to the
findings of N-body simulations. In fact,
the fundamental reason why LP and FF cannot produce large scale
filaments is because they never use power on large scales and
hence cannot build up matter coherently on such scales.

TZ predicts a much lower value of the statistic
on all scales and at all epochs. One may think this contrary
to what the figures \ref {visual}a-d seem to indicate,
especially on very large scales.  However, we recall that in
TZ simulations the amount of matter in overdense
regions is under 30\% at most times
(cf. Fig. \ref {clumps}) indicating
that a bulk  of the matter is distributed diffusely in underdense
regions. This
means that the strong visual signal that we see (especially
in Fig.\ref{visual}c) is not picked up by this statistical test.
Consequently, we obtain, in this case, a very low value of $S.$
(One might anticipate a better agreement between TZ and N-body if the low
signal-to-noise ratio in TZ were amplified using the Minimal Spanning Tree
or a fixed overdensity threshold on which to test for filamentarity.)

\subsubsection {Evolution of voids} \label {ev}

The next in our list of structural units are voids. There is
now a consensus on the view that most of the universe
is filled with voids of different shapes and sizes,
with most of the matter residing on boundaries separating them. The
distribution of volume amongst voids of various sizes, sometimes
called the ``void spectrum'',
is quite sensitive to the primordial spectrum of density
fluctuations \cite{km92,sss94}. Hence a proper understanding of the
distribution of voids in the Universe could in principle
lead to a precise estimation of the primordial spectrum.
Further, Sahni, Sathyaprakash and Shandarin 1994, have pointed
out that the void spectrum can potentially be used to determine the
value of the density parameter $\Omega$ if the shape and
amplitude of the initial perturbation spectrum are independently
known. Thus, the statistics of voids is an important measure
in characterizing the large scale structure of the Universe.
In our study of voids we employ two indicators of void statistics:
(i) The total number of voids in our simulations (which is a function of
epoch),
and (ii) the void
probability function. The former will be discussed here
and the latter in the next subsection.

We define individual voids as
connected regions of a given underdensity. As in the case
of clumps the number of voids is sensitive to the threshold
density chosen for identifying them. Again, since the aim
here is to compare the various approximation schemes with N-body,
we need only choose an appropriate threshold density so that
voids are (visually) well defined. We found that in order
to obtain a good picture of voids it is necessary to smoothen
the density field before applying the density threshold in
selecting void regions. If the density field were not smoothed
then there would be too many tiny voids which would give rise to
a lot of ``noise'' in the evolution of the number of voids
without at the same time making any significant contribution
to the total volume occupied by voids. In Fig. \ref {underdensity}
we have shown regions in our N-body simulations of density
$\rho \le \rho_c$ at two different epochs each
for the $n=0,$ $k_c=32k_f$ model (top panels) and
$n=0,$ $k_c= 256 k_f$ model (bottom panels).
The density fields were smoothed by removing power on
modes $k\ge 64 k_f.$ We see that voids are well defined for
the chosen underdensity except that there are still a few
voids of very small size. Thus, in our definition of voids
we do not take into account voids whose diameters are smaller than
10 grid units.

The evolution of the number of voids plotted in Fig. \ref {voids}
exhibits behaviour similar to the evolution of
clumps seen in Fig. \ref {clumps}: With reference to the
N-body curve (thick solid line) we observe that initially
there is a sharp increase in the number of voids, it acquires
a maximum sometime after the epoch $\sigma_*$
(except in the $n=-2$ model where there is no peak),
and thereafter falls steadily, more or
less stabilizing after a while. The voids are not well defined at
very early epochs ($\sigma \le 1$) but by the epoch $\sigma_*$
they gain their identity. The decline in the number of voids
in Fig. \ref {voids} is a consequence of the fact that voids compete for
space during expansion, and that smaller voids can be
encroached upon by larger ones. Thus, voids not only expand
they can also contract and ultimately disappear \cite{sss94}.

The adhesion model produces roughly the right evolution for the
number of voids, the largest disagreement being for the
$n=0,$ $k_c=256 k_f$ model. In
the latter case it predicts a slightly smaller number of voids
than what is predicted by N-body simulations.
FF and LP fail to reproduce the
correct evolution of the number voids in all but the $n=0,$
$k_c=256 k_f$ model where they agree with N-body very well.
In the case of FF as matter falls into deeper wells of the
potential, the cellular structure gets completely phased
out with the result that at later epochs very few
voids are left behind in this case.  In other words, in FF there
is no cellular structure to ``support'' the voids. The same is
true in the case of LP but because of the fact that the particles do
cross over caustic regions the cellular structure lasts a little
longer and the fall off in the number of voids is somewhat slower than
in the case of FF. Note especially that in the $n=2$ model
both FF and LP predict only one void at all epochs. This is
consistent with the visual pictures corresponding to these
approximations (cf. Fig. \ref {visual}c) which indicate that
these simulations will only have one void with a sponge like
topology. TZ, as expected from the
pictures in Fig. \ref {visual}, always predicts
fewer but larger voids as compared to N-body. This is because regions that are
populated by many small clumps in N--body are filled with a low rather uniform
density in TZ.
It is therefore not a suitable tool to
study the void spectrum.

Summarising, the adhesion model is best suited for studying
the statistics of voids:
It not only predicts the right evolution for the number of voids
(the sharp rise and the subsequent gradual fall-off ),
it also predicts the right
number of voids for most spectra at virtually all times.

\subsubsection {Void probability function} \label {evpf}

At late times voids are large scale coherent features and it
is hard to draw a definite demarcation boundary about a void.
There is great danger in using the ``friends-of-friends" algorithm
to identify voids, especially at late times. A closer look at
Fig. \ref {underdensity} shows that a tiny bridge connecting
two neighbouring voids will cause the algorithm to declare as one
void what visually would appear to be two distinct voids.
For such epochs the void probability function (henceforth referred
to as VPF) is a better indicator of the sizes of voids
than void number. We therefore supplement
the information obtained by studying the evolution of the
number of voids with that obtained using the VPF. The VPF describes
the probability that a sphere of size $R$ (a circle in 2D)
thrown at random is
completely devoid of matter \cite {w79}. In practice one can relax this
condition
a little and say that the VPF is the probability of finding that a
sphere of radius $R$ placed at random within the simulation
box is an ``underdense'' region. Here again we are faced with a
non-objective definition since the results will depend upon
the chosen value of underdensity.  However, for our purposes of comparison,
it makes sense to choose any reasonable underdensity that
would be consistent with what the pictures project.
We first make a map of the overdense regions
as in, say, Fig. \ref {overdensity}  and for this map we compute
the VPF. We have chosen at random 20 \% of all the
grid points in computing the VPF.

Our results are shown in Fig. \ref {vpf}a and b (two
epochs each) for all the models considered by us.
The epochs chosen are the same as in Fig. \ref {fil}.
We notice that the VPF at earlier epochs falls off sharply
with scale indicating a scarcity of large voids. At later epochs
however, the fall off is slower, indicating the formation of
larger voids as the Universe expands.
TZ predicts, as expected, a lot more voids
on all scales and at all epochs as compared to N-body. The
reason for this is simple: At any epoch $\sigma$ all the power
in the primordial potential below the scale $k_{\rm NL} (\sigma)^{-1}$
has been removed in evolving the particles up to the epoch
$\sigma$ and hence one cannot expect the formation of structure
on this and lower scales in TZ simulations. In the $n=0$ models,
at earlier epochs AM, FF and LP, all agree with N-body.
However, at later epochs
the latter two approximations predict fewer voids
of all sizes. The reason for this is that matter does not
participate in hierarchical clustering in FF and LP and thus
matter does not get emptied out as happens in the
case of N-body. AM comes closest to the predictions of N-body for this
spectrum.

While this is the story in the case of spectra
with equal power on all scales the results are totally different
when the power is not equal on all scales. In the case of the
$n=2$ model none of the approximations agree with N-body
with AM and TZ being closest in accuracy.
For $n=-2$ on the other hand FF and LP
agree remarkably well with the predictions of N-body both
at earlier and later epochs and on almost all scales.

\section {Conclusions}

In this work we have studied gravitational instability in the strongly
non-linear regime using a number of distinct and sometimes orthogonal
statistical
indicators such as: Correlation coefficient of particle positions/densities;
statistics of overdense and underdense regions (clumps and voids);
filamentary statistics; and the
Void probability function. Using these statistics we assess the accuracy
of different approximations to gravitational instability such as:
the Truncated Zel'dovich approximation (TZ), the adhesion model (AM),
the frozen flow (FF) and the linear potential (LP) approximation.
We compare these approximations with N-body simulations for a variety of
spectra and at different cosmological epochs.
We find that as the scale of non-linearity grows, so do the characteristic
features of structure such as filamentarity and the sizes of voids.
Our study shows that as long as the non-linear length scale
$k_{\rm NL}^{-1}$ remains smaller than $R_*$ -- the typical distance between
peaks of the gravitational
potential -- all approximations give results in reasonable agreement
with those of N-body simulations. During the epoch $R_* < k_{\rm NL}^{-1} <
R_{\varphi}$ ($R_{\varphi}$ being the correlation length of the potential)
the adhesion
model (and occasionally TZ) gives results closest to N-body.
The reason for this is the following:
Particles in FF follow the streamlines of the initial velocity field,
converging in the minima of the gravitational potential. As a result,
filaments in this approximation, show a tendency to thin out and ultimately
disappear. In the case of LP, particles begin to oscillate about the minima
of the potential at late times, freezing the possibility of any long range
dynamics. Consequently, no real evolution of particle positions can take place
beyond the epoch $k_{\rm NL}^{-1} \simeq R_*$ in either FF or LP which break
down
when $k_{\rm NL}^{-1} > R_*$.

Unlike FF and LP both AM and TZ use the presence of long range modes in
the gravitational potential to move particles at late times. Consequently,
as long as the potential has sufficient long range power
to affect bulk motion over large scales (described by $R_{\varphi}$),
TZ and AM give results closely matching with those of N-body. Compared with
AM, TZ suffers from one major drawback: although it does manage to collect the
right amount of matter into large clumps it completely overlooks the presence
of small clumps. The reason for this is clear, in order to prevent shell
crossing all modes which have gone non-linear by a given epoch are surgically
removed from the initial gravitational potential because of which no small
scale clustering is present in TZ. The adhesion model does not suffer
from this drawback and can accurately predict the multiplicity function
of clumps and the void spectrum as long as $k_{\rm NL}^{-1} \le R_{\varphi}$.
However, after $k_{\rm NL}^{-1} \ge R_{\varphi}$, TZ makes much more
accurate predictions about the location of mass.  In this regime, AM
continues to make better statistical predictions, but its dynamical
accuracy is reduced. This can be understood as inaccurate influence
of the short modes
on the position of structure in AM at late times
especially since
by construction, the adhesion technique when
applied to truncated initial spectrum closely reproduces TZ.

A comparison of $R_*$ and $R_{\varphi}$ for different spectra
(Fig. \ref {gamma}), shows that $R_* \simeq R_{\varphi}$ for very steep or very
shallow
spectra with $n > 2$ or $n < -2$. For intermediate values $-2 \le n \le 2$
$R_{\varphi} \gg R_*$ indicating that for such spectra the adhesion model
(and occasionally TZ)
will be more accurate than FF or LP. These values of the
two dimensional spectral index correspond in three dimensions to the
range $-3 \le n \le 1$ which is precisely the range of interest in most
cosmological scenario's such as CDM. (For the standard CDM model, $R_* < 1$
Mpc. $R_{\varphi} \simeq 50$ Mpc.) We therefore feel that the adhesion
model and the truncated Zel'dovich approximation (depending on which aspect
of the description of structure one wishes to emphasize) are
more realistic approximations to apply to the study of large scale structure
than either FF or LP.

\noindent {\bf Acknowledgments:} ALM wishes to thank the following
(all in the USA) for support: The National Center for Supercomputing
Applications at Urbana, Illinois, NASA (grant NAGW--3832), and NSF (grant
AST--9021414). ALM, DP and VS are grateful to the Aspen (Colorado, USA)
Center for
Physics for sponsoring a 1994 workshop on this and related topics where
this work was completed.

\vfill\eject

\begin{figure}
\caption {The two natural scales of the potential $R_*$ and $R_\varphi$ (left
panel)
and the ``temperature'' $\gamma$ (right panel) are shown
plotted against the spectral index $n$ for featureless power-law spectra.
The values of $R_*$ and $R_\varphi$ are quoted for a box of size
$512\times 512.$  Note that the difference between
the two scales decreases as $|n|$ increases, being largest when $n\simeq 0$
($R_*$ and $R_\varphi$ characterize the mean distance between peaks
and the correlation length of the potential, respectively.)}
\label {gamma}
\end{figure}

\begin{figure}
\caption {Comparison of N-body simulations with the simulations of
various approximation schemes.  Only a fourth of all the particles
are shown. For the sake of clarity, we have superimposed the results
of adhesion model over the particle positions of N-body simulations.
 From top to bottom the pictures correspond to N-body, adhesion model,
frozen flow, linear potential, and truncated Zel'dovich approximation,
respectively. The left and the right panels are obtained
at epochs $\sigma_1$ and $\sigma_2,$ respectively.
The plots are shown for
(a) $n=0,$ $k_c = 32 k_f,$ $\sigma_1=4,$ $\sigma_2=16,$
(b) $n=0,$ $k_c = 256 k_f,$ $\sigma_1=32,$ $\sigma_2=128,$
(c) $n=2,$ $k_c = 256 k_f,$ $\sigma_1=256,$ $\sigma_2=1024,$and
(d) $n=-2,$ $k_c = 256 k_f,$ $\sigma_1=2.00,$ $\sigma_2=2.83.$}
\label {visual}
\end{figure}
\begin{figure}
\caption {Contour plots of the initial potentials corresponding to
the different initial potentials used in our simulations. The solid
lines correspond to $\varphi > 0$ and the dashed lines
to $\varphi < 0.$}
\label {contour}
\end{figure}

\begin{figure}
\caption {Time evolution of the vector correlation coefficient of
particle positions corresponding to
(i) frozen flow (dashed line),
(ii) linear potential (dotted line), and
(iii) truncated Zel'dovich (dashed-dotted line).
Top left panel corresponds to $n=0,$ $k_c=32 k_f$ power law model
and top right panel to $n=0,$ $k_c= 256 k_f$ power law model.
Bottom left panel is for $n=2,$ $k_c=32k_f$ and the one on the
bottom right is for $n=-2,$ $k_c=256k_f.$
The dotted and the solid vertical lines correspond to the epochs
when the scales going nonlinear are $R_*$ and $R_\varphi,$ respectively.}
\label {vccfig}
\end{figure}

\begin{figure}
\caption {Evolution of the density correlation coefficient
corresponding to
(i) adhesion model (solid line),
(ii) frozen flow (dashed line),
(iii) linear potential (dotted line), and
(iv)  Zeldovich (dashed-dotted line).
In Fig.5a the panels are arranged as in Fig. 4 and before the correlation
coefficient is computed the $n=0,$ $k_c=32 k_f$ density
field in each simulation is smoothed at $k_G=16 k_f$ and the rest
of the density fields are smoothed at $k_G=64 k_f.$ In Fig.5b
we have shown the evolution of the density correlation coefficient,
for the $n=0,$ $k_c=256k_f$ model, obtained by smoothing the density
fields, at a given epcoh, at a scale proportional to the nonlinear
scale at that epoch : $k_G^{-1} = 0.5 \times k_{\rm NL}^{-1}.$
This demonstrates that
while the TZ reproduces the large scale features very accurately,
the other approximations, with the exception of the AM, even after such
a smoothing, do not show good agreement with N-body results.}
\label {sccfig}
\end{figure}

\begin{figure}
\caption {Plot showing regions of density greater than the threshold
density $\rho_c$ in our N-body simulation for $n=0$ models.
Notice that for the chosen threshold density clumps are well defined
features. Top panels correspond to $k_c=32 k_f$ and bottom panels to
$k_c= 256 k_f.$}
\label {overdensity}
\end{figure}

\begin{figure}
\caption {Evolution of the number of clumps (top panels) and the total
mass in clumps (bottom panels) corresponding to
(i) N-body  simulations (thick solid line),
(ii) adhesion model (solid line),
(iii) frozen flow (dashed line),
(iv) linear potential (dotted line), and
(v) truncated Zel'dovich (dashed-dotted line).
The evolution is shown for: (a) $n=0$
spectrum with left and right panels corresponding, respectively,
to, $k_c=32k_f$ and $k_c=256k_f,$ and (b) for
$n=2,$ $k_c= 256k_f$ spectrum (left panels) and
$n=-2,$ $k_c= 256k_f$ spectrum (right panels).}
\label {clumps}
\end{figure}

\begin{figure}
\caption {Filamentary statistic $S(R)$ is shown plotted as
a function of the scale length $R$ for two different epochs
(as indicated by the value of $\sigma$ within the panels)
corresponding to
(i) N-body (thick solid line),
(ii) frozen flow (dashed line),
(iii) linear potential (dotted line), and
(iv) truncated Zel'dovich (dashed-dotted line).
Results are shown for (a) $n=0$ with $k_c = 32 k_f$ (top panels),
$k_ c= 256 k_f$ (bottom panels) and
(b) $n=2,$ $k_c= 256k_f$ spectrum (top panels) and
$n=-2,$ $k_c= 256k_f$ spectrum (bottom panels).}
\label {fil}
\end{figure}

\begin{figure}
\caption {Plot showing regions in our N-body simulations
of density less than a threshold density $\rho_c$
for $n=0$ models at two epochs each for
cutoff at $k_c=32k_f$ (top panels) and $k_c = 256 k_f$
(bottom panels). }
\label {underdensity}
\end{figure}

\begin{figure}
\caption {Evolution of the number of voids corresponding to
(i) N-body  simulations (thick solid line),
(ii) adhesion model (solid line),
(iii) frozen flow (dashed line),
(iv) linear potential (dotted line), and
(v) truncated Zel'dovich (dashed-dotted line).
The panels are as in Fig. 4.}
\label {voids}
\end{figure}

\begin{figure}
\caption {Void probability function $V(R)$ is shown
plotted as a function of the scale $R$ for two different epochs
(as indicated by the value of $\sigma$ within the panels)
corresponding to:
(i) N-body  simulations (thick solid line),
(ii) adhesion model (solid line),
(iii) frozen flow (dashed line),
(iv) linear potential (dotted line), and
(v) truncated Zel'dovich (dashed-dotted line).
Results are shown for (a) $n=0$ with $k_c = 32 k_f$ (top panel),
$k_ c= 256 k_f$ (bottom panel) and
(b) $n=2,$ $k_c= 256k_f$ spectrum (top panel) and
$n=-2,$ $k_c= 256k_f$ spectrum (bottom panel).
The overdensity field needed in finding VPF
is determined by taking a threshold density
of $\rho_c=5\rho_0$ in all but $n=0,$ $k_c=256k_f$ model for
which the threshold is taken to be $\rho_c=2\rho_0.$}
\label {vpf}
\end{figure}

\end{document}